\documentclass[reprint,aps,prl,floats]{revtex4-1}

\usepackage[dvips]{color}
\usepackage{graphicx}
\usepackage{bm}
\usepackage{textcomp} 
\usepackage{amsmath} 

\newcommand{\be}{\begin{equation}}
\newcommand{\ee}{\end{equation}}
\newcommand{\ud}{d}
\newcommand{\eps}{\varepsilon}

\begin{document}

\title{Universal lineshapes at the crossover between weak and strong critical coupling in Fano-resonant coupled oscillators}

\author{Simone Zanotto}
\email{simone.zanotto@alumni.sns.it}
\affiliation{NEST, Istituto Nanoscienze - CNR and Scuola Normale Superiore, P.za S. Silvestro 12, 56127 Pisa, Italy}

\author{Alessandro Tredicucci}
\affiliation{NEST, Istituto Nanoscienze - CNR, and Dipartimento di Fisica ``E. Fermi'', Universit\`a di Pisa, Largo Pontecorvo 3, 56127 Pisa, Italy}

\date{\today}

\begin{abstract}
In this article we discuss a model describing key features concerning the lineshapes and the coherent absorption conditions in Fano-resonant dissipative coupled oscillators. The model treats on the same footing the weak and strong coupling regimes, and includes the critical coupling concept, which is of great relevance in numerous applications; in addition, the role of asymmetry is thoroughly analyzed. Due to the wide generality of the model, which can be adapted to various frameworks like nanophotonics, plasmonics, and optomechanics, we envisage that the analytical formulas presented here will be crucial to effectively design devices and to interpret experimental results.  
\end{abstract}

\maketitle

Among the resonance lineshapes, Fano lineshapes deserve a special interest since they first enabled to interpret atomic and molecular physical processes \cite{FanoPR1961}, and subsequently the response of nanostructured systems like photonic crystals and plasmonic resonators \cite{LukyanchukNatMat2010}. In general, resonances may be absorptive, in the sense that part of the energy vehicled by the excitation field is converted into other forms. While absorption, in the sense of losses, is usually an undesired effect, in other frameworks it can be harnessed to enable functional operations like detection, wavelength conversion or quantum state transfer. Absorbing systems also exhibit a rich physics, as for instance that of coherent perfect absorption (CPA) which shares some key mathematical aspects with non-Hermitian quantum systems and parity-time symmetry breaking \cite{LonghiPRA2010, ChongPRL2011, MingPRA2013, YongPRL2014}. In addition, CPA and related concepts may open new avenues in the control of wave properties like polarization \cite{KangArxiv2015} or in the processing of chaotic signals \cite{LonghiPRA2012}.

In its original formulation, the theory of CPA is very general, as it relies on fundamental properties of the scattering matrices \cite{ChongPRL2010}; when dealing with specific systems, appropriate modeling tools are needed. However, due to the possibly complex nature of the systems under analysis, microscopical approaches are often time-consuming, and an analytical model would be an advantage both for the interpretation of experimental results and as a guide to target ab-initio simulations. In this paper we analyze the CPA in a two-oscillator coupled-mode model, which, thanks to its generality, can be applied to a number of emergent frameworks like photon- or plasmon-exciton coupled systems\cite{ArtusoNL2008, ManjavacasNL2011, RidolfoPRA2010, WaksPRA2010, LalannePRX2015} and optomechanics\cite{KenanPRA2013}. It will be first shown that the Fano trasmittance and reflectance resonances typical of a single symmetric resonator are inherited by the coupled asymmetric system. Absorption lineshapes, instead, are described by another, universal lineshape, depending on few, physically meaningful parameters. A check of the model validity is also provided, based on the concept of strong light-matter coupling in a realistic resonant metasurface embedding intersubband-active quantum wells.

\begin{figure}[htbp]
\centerline{\includegraphics{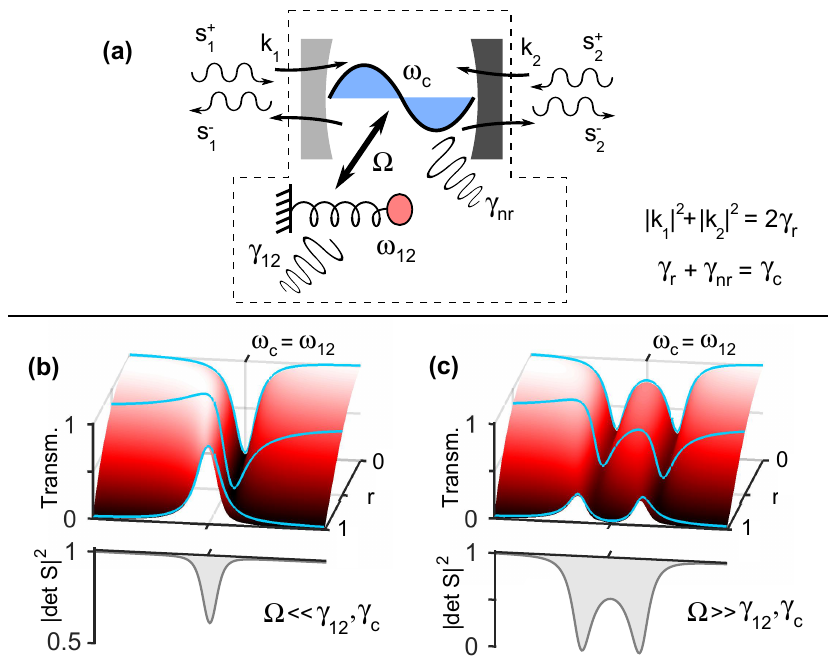}}
\caption{Panel (a): sketch of the coupled oscillator model analyzed in the article. Panels (b-c): spectral lineshapes of the weakly (b) and strongly coupled (c) system. The  transmittance lineshape can be tuned through a parameter ($r$, see text), and is inherited from the weakly to the strongly coupled case. The S-matrix determinant, instead, is given by a universal function, independent of the trasmittance (and reflectance) lineshapes. Parameter values are $\gamma_r = \gamma_{12}$, $\omega_c = \omega_{12} = 50 \gamma_r$, $\gamma_{nr} = 0$. In case (b) the ratio $\Omega/\gamma_r $ equals to 0.3, while in case (c) it equals to 2.7.}
\end{figure}

The model under consideration is schematized in Fig.~1 (a). A resonant cavity at frequency $\omega_c$ is coupled to a second resonant degree of freedom, here represented as a spring-mass resonator at $\omega_{12}$, through a coupling coefficient $\Omega$. From now on, the second oscillator will be referred to as ``matter'' resonator, since a prototypical situation would be that of a two-level system (atom, exciton) treated under the semiclassical approximation. However, another important situation which can be described by the present model is that of a  plasmonic system supporting a ``bright'' and a ``dark'' (subradiant) mode: the oscillator at $\omega_{12}$ here represents the dark mode. The cavity resonator radiates into, and is excited from, two radiative scattering channels, with couplings $\kappa_{1,2}$. $\gamma_{12}$ describes an internal loss mechanism of the matter resonator, while $\gamma_{nr}$ describes a \textit{non-radiative} and \textit{non-resonant} cavity loss mechanism. In the photonic framework, $\gamma_{nr}$ may represent losses such as roughness scattering or dissipation in a metal component.

The dynamics of the system is described by 
\begin{eqnarray}
 \frac{\ud b}{\ud t} & = & (i \omega_{12} - \gamma_{12}) b + i \Omega a \nonumber \\
 \frac{\ud a}{\ud t} & = & (i \omega_{c} - \gamma_{c}) a + i \Omega b + ( \langle \kappa | ^* ) | s^+ \rangle \nonumber \\
 | s^- \rangle       & = & C | s^+ \rangle + a | \kappa \rangle 
\label{cmt_2port}
\end{eqnarray}
where $a$ and $b$ are the amplitudes of, respectively, the cavity and the matter resonators. Here, $| s^\pm \rangle = (s_1^\pm, s_2^\pm)^T$ describe the amplitudes of ingoing and outgoing scattered waves. Similarly, $| \kappa \rangle = (\kappa_1, \kappa_2)^T$. In this notation, $| v \rangle \rightarrow \langle v |$ means transposition and complex conjugation, while $| v \rangle \rightarrow | v \rangle^*$ means only complex conjugation. $C$ is a matrix which describes the non resonant scattering process.

The free evolution of the cavity resonator occurs with a damping rate $\gamma_c$, which describes its total losses, radiative plus nonradiative. It can be decomposed as $\gamma_c = \gamma_r + \gamma_{nr}$, where the second term has the meaning described above, while the first (the purely radiative damping rate) must satisfy $2 \gamma_r = \langle \kappa | \kappa \rangle$. This constraint can be derived by imposing instantaneous energy conservation to Eqs.~\ref{cmt_2port}. Energy conservation and time-reversal symmetry constraints also require $C | \kappa \rangle^* = -| \kappa \rangle$, as already observed for the single-oscillator case \cite{FanJOSAA2003}. 

The system's response is fully described by its scattering matrix $S$, which links the amplitudes of ingoing and outgoing waves through $| s^-_0 \rangle = S(\omega) | s^+_0 \rangle$. Direct integration of Eqs.~\ref{cmt_2port} yields
\be
S(\omega) = C - \frac{i(\omega-\omega_{12})+\gamma_{12}}{(\omega - \omega_+)(\omega - \omega_-)} D
\label{cmt_s}
\ee
where $D = | \kappa \rangle \left( \langle \kappa |^* \right) $. The explicit expression of the poles $\omega_{\pm}$ will be given in the following, while those of matrices $C$ and $D$ are given in the Supplementary Material. 

Matrix $C$ depends on a parameter $r \in [0,1]$, which describes the off-resonant reflection amplitude. Matrix $D$ involves an additional parameter $\xi$, which describes the asymmetry of the decay into the scattering channels. This parameter, constrained in the interval $[-1,1]$, is connected with the coupling coefficients through the relation $r\xi = (|\kappa_1|^2 - |\kappa_2|^2)/(|\kappa_1|^2 + |\kappa_2|^2)$. This link between $\kappa_{1,2}$ and $r$, already outlined for single-mode optical resonators \cite{FanOL2013}, is here generalized to the two-oscillator model.

In this model, if the cavity is decoupled from the matter resonator ($\Omega = 0$) and there are zero non-resonant losses ($\gamma_{nr} = 0$), the results of Ref.~\cite{FanOL2013} are recovered. Considering transmission and reflection spectra, Fano lineshapes are obtained. Similar lineshapes, although less contrasted, are observed under the weak cavity-matter coupling regime, i.e., when the coupling $\Omega$ is finite but smaller than $\gamma_{12}$ and $\gamma_{c}$. As shown in Fig.~1 (b), the lineshapes sweep from a Lorentzian to an inverted Lorentzian according to the value of $r$. Here, we assumed $\xi = 0$, while it can be shown that  $\xi \neq 0$ leads to a further decrased lineshape contrast. Consider now the spectra in Fig.~1 (c), obtained when $\Omega$ is larger than $\gamma_{12}$ and $\gamma_{c}$. The two resonators are strongly coupled, and the spectral feature is doubled, following the peak, dip, or asymmetric shape of the weakly coupled system. This unique behaviour (``\textit{lineshape inheritance}''), which was already observed experimentally and justified heuristically \cite{ZanottoPRB2012}, is now grounded on a basic theoretical model, and can be extended to all systems which can be described by Eqs.~(1). 

In the above, the focus was on transmittance and reflectance. While in an experiment these are the most easily accessible quantities, a more meaningful probe of a driven linear system would rather be the absorption, since it is directly connected to the excitation of the resonant degrees of freedom. Here, the model under consideration presents two scattering channels, and should be analyzed in view of the coherent absorption theory for asymmetric two-port systems \cite{ChongPRL2010, BaldacciOE2015}. A key quantity is the $S$-matrix determinant, which, following Eq.~\ref{cmt_s}, can be expressed as $ \det{S(\omega)} = e^{2 i \phi} (\omega - \bar{\omega}_+)(\omega - \bar{\omega}_-)/(\omega - \omega_+)(\omega - \omega_-) $. $\phi$ is a global phase (see Supplementary Materials), while the zeroes, which are connected to the coherent perfect absorption (CPA), explicitly read
\begin{equation}
\begin{split}
\bar{\omega}_{\pm} & = \frac{\omega_c + \omega_{12}}{2} -  i \frac{\gamma_r - \gamma_{nr} - \gamma_{12}}{2} \\ &\quad \pm \frac{1}{2}\sqrt{ \left[(\omega_c - \omega_{12} )  - i (\gamma_r - \gamma_{nr} + \gamma_{12}) \right]^2 + 4 \Omega^2 }. 
\end{split}
\end{equation}
The poles $\omega_{\pm}$ are obtained from the zeroes by replacing $\gamma_r$ with $-\gamma_r$. The essential feature is that $\det{S}$ depends \textit{neither} on $r$ \textit{nor} on $\xi$: the lineshape-governing factors do not influence the $S$-matrix determinant. Nor they influence the CPA condition, too: it can be shown that $\det{S} = 0$ if and only if
\be
\gamma_{-}^{2} = \frac{1}{2} \left[ \gamma_{+}^2 - \delta^2 - 4 \Omega^2  + 
\sqrt{(\gamma_{+}^2 - \delta^2 - 4 \Omega^2)^2 + 4 \delta^2 \gamma_{+}^2 } \right]
\label{cmt_CPA}
\ee 
where $\gamma_{-} = \gamma_r - \gamma_{nr} - \gamma_{12}$, $\gamma_{+} = \gamma_r - \gamma_{nr} + \gamma_{12}$,  and $\delta = \omega_{12} - \omega_{c}$. 

Eq.~\ref{cmt_CPA} unifies and generalizes the weak and strong critical coupling concepts (WCC and SCC), which were introduced in Ref.~\cite{ZanottoNatPhys2014} for a symmetric ($\xi = 0$) and degenerate ($\delta = 0$) coupled resonator system. As recalled in Fig.~2(a), the WCC and SCC regimes appeared there to be well separated curves on the system's phase diagram, merging into a single exceptional point. 

\begin{figure}[htbp]
\centerline{\includegraphics{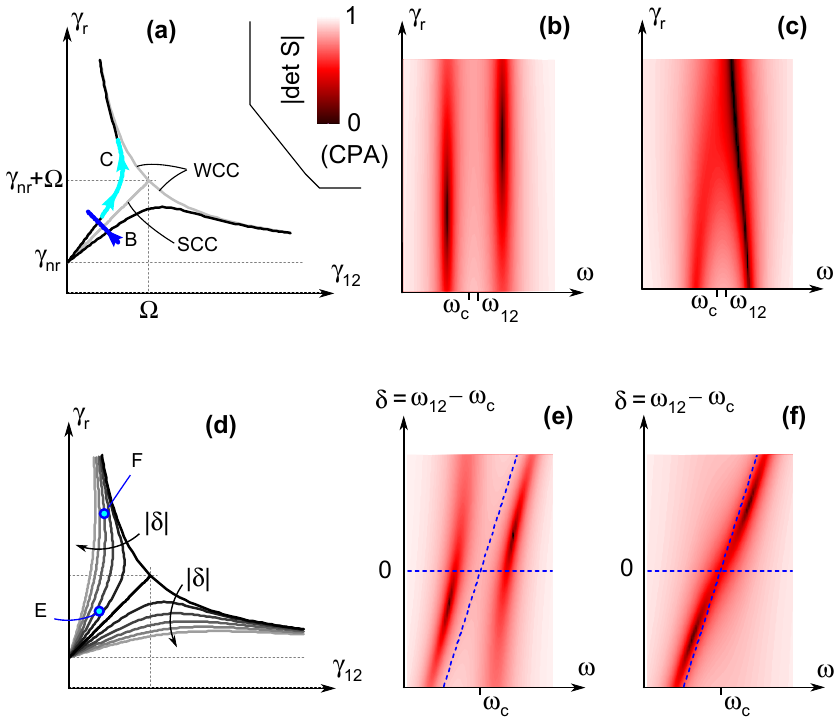}}
\caption{Coherent perfect absorption (CPA) in the non-degenerate coupled oscillator model. The panels on the left represent the phase diagram of the system: on the thick curves, CPA occurs. Well separated weak and strong critical coupling (WCC and SCC), encountered in the degenerate case ($\delta = 0$), are no more distinct in the $\delta \neq 0$ case. Panels (b) and (c) represent the S-matrix spectra for the parametric paths labeled B and C in panel (a), for a fixed value of $\delta$. Similarly, panels (e) and (f) correspond to points labeled E and F in panel (d), where instead $\delta$ is swept.}
\end{figure}

When instead $\delta \neq 0$, the solution of Eq.~\ref{cmt_CPA} are represented by the thick lines in Fig.~2(a). Hence, for the non-degenerate case, well separated WCC and SCC do not exist anymore, and a smooth transition is instead observed. However, some features reminiscent of WCC and SCC are still evident when the $\det S$ spectra are considered. These spectra are reported in panels (b-c), which correspond to the parametric paths labeled B and C in panel (a). In case B, $\det S$ has still a double-dip, but, as opposed to the $\delta = 0$ case, CPA never occurs simultaneously for two frequencies. When the transition between the SCC and WCC region is explored (path C), the coalescence of $|\det{S}\ |$ minima is observed, but only the higher frequency dip is a CPA. (Whether the CPA originates from the dip at higher frequency, or from that at lower, is determined by the signs of $\delta$ and of $\gamma_{-}$). The effect of a continuous sweep of the detuning $\delta$ is analyzed in Fig.~2 (d--f). Suppose that the system is close to, but not exactly on, the SCC (point E). Its spectrum, if $\delta = 0$, has two dips which do not reach zero, as highlighted by the dashed horizontal line in (e); if $\delta$ is tuned, a CPA occurs either on the lower or on the higher frequency resonance of the coupled system. Suppose instead that the system is close to the WCC (point F). Again, by tuning $\delta$ the system can be brought to CPA, but now with a single isolated spectral feature. In essence, the detuning has a twofold role. On one hand, it weakens the distinction between strong and weak critical coupling, as when $\delta \neq 0$ the curves describing CPA on the phase diagram are smooth and do not exhibit any exceptional point. On the other hand, a proper tuning of $\delta$ can help a system to reach CPA, without acting either on the coupling or on the damping rates. All these observations apply independently of the system's asymmetry and of the specific Fano lineshape observed in reflection/transmission, leading to a universal behaviour of coupled dissipative resonators. 

Nonetheless, the asymmetry plays a singular role in the response of coupled dissipative oscillators. This will be explored by analyzing a realistic structure, a resonant metasurface embedding semiconductor quantum wells (QWs). Such device implements a prototypical system in which weak and strong coupling have been observed, and where they can be harnessed to develop efficient mid-infrared and terahertz light sources, as well as functional optical components \cite{ZanottoPRB2015, ManceauPRX2015}.    

A schematic of the device is sketched in Fig.~3 (a). It consists of a heterostructured membrane of 60 GaAs/Al$_{0.33}$Ga$_{0.67}$As QWs, with well/barrier thicknesses 6.8/20 nm resulting in an intersubband transition frequency $\hbar \omega_{12} \simeq 150\ \mathrm{meV}$. The membrane has thickness $t_2 = 1.3\ \mu \mathrm{m}$, and is periodically patterned with 50 nm thick gold stripes, whose spacing is $a \simeq 3.5\ \mu \mathrm{m}$ and filling fraction is $f \simeq 0.8$. A high-index coating ($\eps = 10$) with thickness $t_1$ completes the stack. 
\begin{figure}[ht]
\centerline{\includegraphics{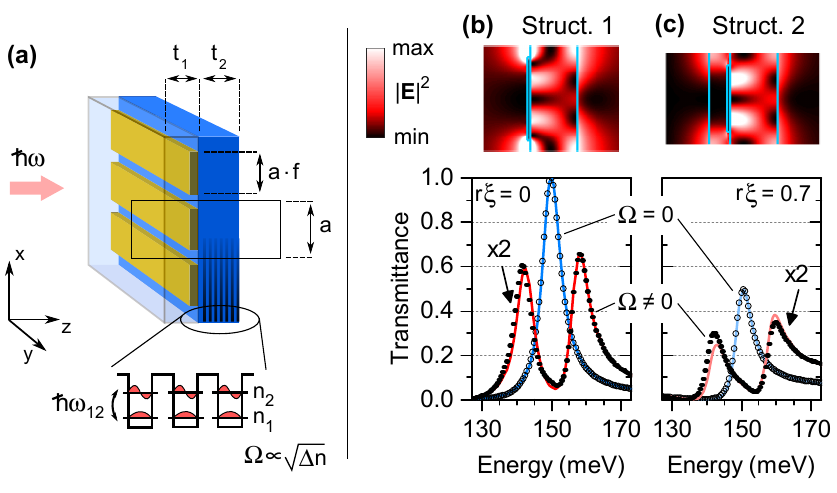}}
\caption{Panel (a), schematic of a resonant metasurface embedding quantum wells, which implement a prototype of strongly and critically coupled oscillators with asymmetry. Panels (b) and (c), resonant field and spectral transmittance of two structures which differ by the value of $t_1$. These resonators are described by different asymmetry parameters, resulting in differently contrasted lineshapes. The calculations from a rigorous electromagnetic solver (dots) are faithfully reproduced by the coupled-mode model (red and blue lines).}
\end{figure}
The structure is modeled through rigorous coupled wave analysis (RCWA), following the details given in \cite{ZanottoPRB2012, TodorovPRB2012} and in the Supplementary Material. If the response of the QWs is turned off ($\Omega \propto \sqrt{\Delta n} = 0$), the device exhibits an isolated photonic resonance at $\omega_c \simeq$ 150 meV (dotted traces in Fig.~(3)). Two structures have been analyzed, and labeled 1 and 2; their details are given in the Supplementary Material. Their main difference is in the value of $t_1$, which governs the cavity asymmetry. The RCWA transmittance spectra have been fitted through Eq.~\ref{cmt_s}, which led to $r\xi \simeq 0$ for struct.~1 and $r\xi \simeq 0.7$ for struct.~2. A signature of such non-zero value is the transmittance contrast, which, for struct.~2, is less than unity. When the QW response is turned on ($\Delta n \neq 0$), the transmittance peak splits into two, with the line shape and contrast being inherited. It should be noticed that, for the $\Omega \neq 0$ case, the agreement between the RCWA spectra (dots) and those obtained from the coupled mode model (lines) relies on a first-principle relation which connects $\Delta n$ and $\Omega$ (see Supplementary Material for details).

The most interesting physics occurs however when the effects of the asymmetry on the joint absorption  $A_{\mathrm{j;\pm}}$ of the two-sided cavity are analyzed \cite{BaldacciOE2015}. Absorption of optical radiation from such systems is a meaningful figure of merit, for instance for detectors and for other devices where an efficient pumping of the polariton population is needed. From Eq.~\ref{cmt_s}, one gets 
\begin{equation}
 A_{\mathrm{j;\pm}}(\omega, x) = \frac{ A_{\mathrm{uni}}(\omega) }{2} \left( 1 + r \xi x \pm \sqrt{(1-r^2 \xi^2)(1- x^2)} \right).
\label{eq_ajoint_cmt}
\end{equation}
Here, $x = (|s^{+}_1|^2 - |s^{+}_2|^2)/(|s^{+}_1|^2 + |s^{+}_2|^2)$ is the asymmetry between the input intensities, and the sign $\pm$ represents the minimum and maximum absorption achievable by acting on the relative phase of the input beams. Notice that $A_{\mathrm{j}}$ as a function of $x$ shows the peculiar elliptical behaviour of coherent absorption \cite{BaldacciOE2015}. The significant feature standing out from Eq.~\ref{eq_ajoint_cmt} is that the ellipse is described by the sole function in parentheses, which factors out from the universal spectral lineshape $A_{\mathrm{uni}}(\omega) = 1 - |\det S (\omega)|^2$. Hence, it is only the function in parentheses, which we label $f_{\pm}(r, \xi, x)$,  that summarizes the effect of asymmetry on coherent absorption. Instead, $A_{\mathrm{uni}}$ is independent on both $r$ and $\xi$. This observation goes beyond what has been stated when discussing Eq.~\ref{cmt_CPA}: it is not only the CPA condition which does not involve $r$ and $\xi$, but the joint absorption lineshape in its fullness.

\begin{figure}[htbp]
\centerline{\includegraphics{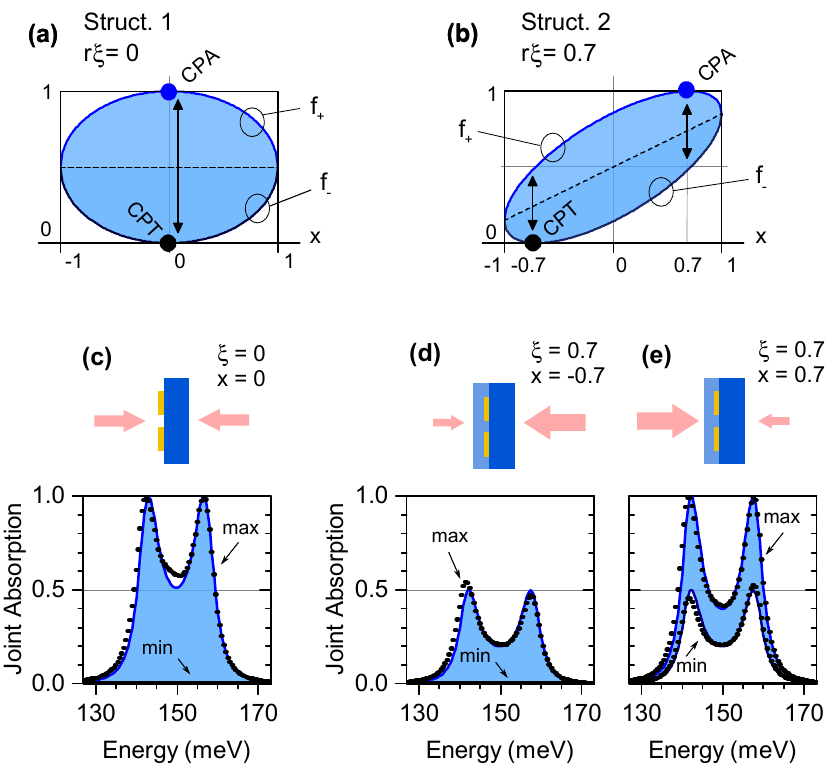}}
\caption{Coherent absorption universal factor $f_{\pm}$ for a  coupled-resonator system which behaves symmetrically (a) or asymmetrically (b). Minimum and maximum joint absorption for structure 1 (c) and structure 2 (d, e). Both structures show coherent perfect absorption and transparency (CPA and CPT), but in the latter those are observed at different $x$-values, i.e., for different states of external excitation, due to the asymmetric behaviour of the cavity.}
\end{figure}
In Fig.~4 (a) and (b) we plot $f_{\pm}(x)$, where the choices $r \xi = 0$ and $r \xi = 0.7$ recall the actual values of structures 1 and 2. The function $f_{-}$ is zero for $x = - r \xi$, which means that the system can always exhibit coherent perfect transparency (CPT) provided that the excitation intensities properly match the intrinsic asymmetry of the photonic resonance. For what concerns CPA, instead, necessary and sufficient condition is that $A_{\mathrm{uni}} = 1$ (and hence $|\det S| = 0$), \textit{and} $f_{+} = 1$, i.e., $x = r \xi$. In other words, there are two independent requirements: the first on damping rates, detuning, and coupling coefficient (Eq.~\ref{cmt_CPA}), the second on the symmetry.  

A numerical test is proposed in Fig.~4 (c-e), where $A_{\mathrm{j}}$ calculated numerically for structures 1 and 2 is compared with the prediction of Eq.~\ref{eq_ajoint_cmt}. No further parameters in addition to those which led to the fitting of Fig.~3 are involved. Since the samples have been designed in order to have $\delta = 0$ and $\gamma_r = \gamma_{12}$, in both cases strong critical coupling occurs and CPA is expected. Indeed, when the proper $x$-value is chosen, $A_{\mathrm{j}}$ reaches 1 in a double-peaked fashion for both structures (Fig.~4 (c) and (e)). Since structure 1 behaves symmetrically ($\xi = 0$), CPA and CPT occur simultaneously for $x = 0$; in structure 2, instead, CPA and CPT occur for opposite values $\xi = \pm 0.7$, consistently with Eq.~\ref{eq_ajoint_cmt}.  

In summary, we studied the absorption lineshapes occurring at the transition between weak and strong critical coupling for a system consisting of two coupled detuned resonators, one of which is radiatively coupled with the exterior in an asymmetric fashion. From this model a peculiar fingerprint in the absorption spectra stands out: a universal lineshape, independent of the asymmetry degree, which instead rules the Fano lineshapes observed in transmittance or in reflectance. Similarly, the coherent perfect absorption (CPA) condition results to be independent of the lineshape-governing factors. Rather, it turns out that the CPA phase diagram is significantly affected by the oscillator's detuning, whose effect is to weaken the distinction between weak and strong critical coupling regimes. Being the present model of wide generality, it is of significance in the development of many active research areas like nanophotonics, plasmonics or optomechanics, where the modeling of a complex system would draw advantage from analytical expressions depending on few parameters of direct interpretation.

\end{document}


\title{Supplementary Materials for the manuscript ``Universal lineshapes at the crossover between weak and strong critical coupling in Fano-resonant coupled oscillators''}

\author{Simone Zanotto}
\email{simone.zanotto@alumni.sns.it}
\affiliation{NEST, Istituto Nanoscienze - CNR and Scuola Normale Superiore, P.za S. Silvestro 12, 56127 Pisa, Italy}

\author{Alessandro Tredicucci}
\affiliation{NEST, Istituto Nanoscienze - CNR, and Dipartimento di Fisica ``E. Fermi'', Universit\`a di Pisa, Largo Pontecorvo 3, 56127 Pisa, Italy}

\date{\today}

\begin{abstract}
These Supplementary Materials contains explicit expressions for certain quantities introduced in the main text, as well as detailed information about the analyzed nanostructured devices.
\end{abstract}

\maketitle

\section{Explicit expressions of matrices $C$ and $D$}
The coupled-mode model presented in the first part of the main article involves two matrices, $C$ and $D$, which represent the non-resonant and the resonant scattering processes, respectively. The elements in these matrices are constrained by energy conservation, time-reversal symmetry and reciprocity. Under these hypotheses, the most general expressions explicitly read 
\[
C = e^{i \phi}\left( \begin{array}{cc}
     r e^{i \psi}  & it \\
     it & r e^{-i \psi} 
    \end{array} \right)
\]
\[
 D = -e^{i \phi} \gamma_{r}\left( \begin{array}{cc}
     d_{11}  & d_{12} \\
     d_{12}  & d_{22} 
    \end{array} \right) 
\]
where
\begin{eqnarray}
d_{11} & = &  \left(r+\xi\pm it \sqrt{1-\xi^2}\right) e^{i \psi} \nonumber \\
d_{22} & = &  \left(r-\xi\pm it \sqrt{1-\xi^2}\right) e^{-i \psi} \nonumber \\
d_{12} & = & \pm r \sqrt{1-\xi^2} + it 
\label{cmt_dd}
\end{eqnarray}

The phases $\phi$ and $\psi$ can be chosen at will, while $r$ is a real parameter belonging to $[0,1]$. The parameter $t$ is connected to $r$ via $t = \sqrt{1-r^2}$ ($t \in [0,1]$). The meaning and the bounds on $\xi$ are discussed in the main article. 

\section{Metasurface modeling and parametrization}
The metasurface is modeled by means of rigorous coupled wave analysis. TM polarization (i.e., incident electric field polarized orthogonal to the metal stripes)  and normal incidence is considered. Permittivity of gold is assumed to be $\eps_{\mathrm{Au}} = -4000+300i$, that of GaAs $\eps_{\mathrm{GaAs}} = 10$, while the quantum well's (QWs) response follows a Lorentz oscillator model whose details are given in Refs.~\cite{ZanottoPRB2012, TodorovPRB2012}. There, the key parameters are the resonance frequency $\omega_{12}$, the damping rate $\gamma_{12}$ and the subband surface charge difference $\Delta n$.

First we considered the QWs to be inactive ($\Delta n = 0$), which means that the metasurface resonance is not coupled to the QWs ($\Omega = 0$). We designed two structures, whose geometrical parameters $a$, $f$ and $t_1$ are reported in Table I, as well as the parameters $\omega_c$, $\gamma_r$, $r$ and $\xi$ which describe the response in terms of coupled mode model. The thickness $t_1$ is the most significant structural parameter, since it rules the transmittance contrast which is directly connected to the asymmetry parameter $\xi$. It turns out that $t_1 = 0$ leads to a fully contrasted transmittance lineshape, which corresponds to $\xi = 0$ and hence to equal decay rates in the two scattering channels ($\kappa_1 = \kappa_2$ in the notation of Fig.~1 of the main text). The parameters $a$ and $f$ have been finely tuned in order to obtain resonance frequencies and linewidths as close as possible for the two structures. By fitting the transmittance spectra of the uncoupled metasurface (empty dots in Fig.~3 in the main text), the parameters  $\omega_c$, $\gamma_r$, $r$ and $\xi$ reported in Table I are obtained. 
It can be noticed that both structures have similar radiative lifetimes $\gamma_{r}$ and non-resonant reflection $r$. Moreover, both structures share similar values for the field overlap factors $\Gamma$. This is defined as $\Gamma = \int_{\mathrm{QW}} |E_z|^2 \eps / \int |\vet E|^2 \eps$ (QW stands here for the quantum well region), and will be of relevance in the following.

In a second step we considered the active QWs ($\Delta n = 5 \times 10^{11}\ \mathrm{cm}^{-2}$). In this case, the metasurface resonance is coupled with the QW transition, with a strength quantified by the vacuum Rabi frequency which reads \cite{ZanottoPRB2012}
\[
\Omega = \sqrt{\frac{e^2 \Delta n \Gamma}{  4 \eps_0 \eps_{\mathrm{GaAs}} m^* L_{per}}}
\]
where $\eps_{\mathrm{GaAs}}$ is the well material permittivity, $m^* = 0.067 m_0$ is the conduction subband effective mass, and $L_{per}$ is the QW period thickness. For the analyzed structures, $\hbar \Omega \simeq 8\ \mathrm{meV}$, and the system is in strong coupling. The fingerprint of strong coupling is evident from the double-peak transmittance curves obtained from the RCWA (dots in Fig.3, main text). In the RCWA the QWs are modeled as a Lorentz oscillator with $\hbar \omega_{12} = 150\ \mathrm{meV}$, $\hbar \gamma_{12} = 3\ \mathrm{meV}$. These numbers, together with the other parameters extracted from the empty cavity spectra, are then plugged into the coupled-mode model, resulting in  the transmittance curves reported as red lines in Fig.~3 of the main text. Such double-peaked spectra should be compared with the dotted traces obtained through the RCWA. The very good agreement witnesses that the analytical expressions provided by the coupled-mode model can quantitatively predict the response of an asymmetric two-port photonic system strongly coupled with a matter excitation. 

\begin{table}[h!]
  \caption{Geometric data and coupled-mode model parameters for the structures analyzed in Figs.~3 and 4.}
  \begin{center}
    \begin{tabular}{r||ccc|cccc|c}
    \hline
     & $t_1$ & $a$ & $f$  & $\omega_c$ & $\gamma_r$ & $r$ & $\xi$ & $\Gamma$\\
      \textbf{Struct.} & [$\mu$m] & [$\mu$m] &   & [meV] & [meV] & & & \\ 
    \hline
    \textbf{1} & 0 & 3.70 & 0.80 &  149.60 & 3.25 & 0.996 & 0.001 & 0.47 \\
    \textbf{2} & 0.5 & 3.20 & 0.73  & 149.70 & 2.77 & 0.981 & 0.72 & 0.56 \\ 
    \hline
    \end{tabular}
  \end{center}
\end{table}